\documentclass[twocolumn,showpacs,preprintnumbers,amsmath,amssymb,prb]{revtex4}
\usepackage{graphicx}
\usepackage{xcolor}
\usepackage[english]{babel}
\usepackage[T1]{fontenc}
\usepackage[latin1]{inputenc}
\usepackage{amsmath}
\usepackage{amssymb}
\usepackage{subfigure}

\def\d{\mbox{d}}

\def     \a{\`{a}~}

\definecolor{red}{rgb}{1,0,0}
\begin{document}

\title{Time persistency of floating particle clusters in free-surface turbulence}

\author{
Salvatore Lovecchio$^{1~}$,
Cristian Marchioli$^{1, 2, 3~}$,
and Alfredo Soldati$^{2, 3~}$\footnote{Author to whom correspondence
should be addressed E-mail: soldati@uniud.it~~Phone: +39~(0)432 558020.}
}
\address{
$^{1}$ Dipartimento di Ingegneria Elettrica, Gestionale e Meccanica,
Universit\a degli Studi di Udine, 33100, Udine, Italy \\
$^{2}$ Centro Interdipartimentale di Fluidodinamica e Idraulica,
Universit\a degli Studi di Udine, 33100, Udine, Italy \\
$^{3}$ Dipartimento di Fluidodinamica, CISM, 33100, Udine,Italy
}


\begin{abstract}
{\bf Abstract} We study the dispersion of light particles floating on a flat shear-free
surface of an open channel in which the flow is turbulent.
This configuration mimics the motion of buoyant matter (e.g. phytoplankton,
pollutants or nutrients) in water bodies when surface waves and ripples
are smooth or absent. We perform direct numerical simulation of turbulence coupled
with Lagrangian particle tracking, considering different values of the
shear Reynolds number ($Re_{\tau}=171$ and $509$)
and of the Stokes number ($0.06 < St < 1$ in viscous units).
Results show that particle buoyancy induces clusters that evolve
towards a long-term fractal distribution in a time
much longer than the Lagrangian integral fluid time scale, indicating
that such clusters over-live the surface turbulent structures which
produced them.
We quantify cluster dynamics, crucial when modeling dispersion in
free-surface flow turbulence, via the time evolution of the
cluster correlation dimension.
\end{abstract}

\maketitle

\section{Introduction}
Buoyant particles transported by three-dimensional incompressible turbulence are known to
distribute non-uniformly within the flow [\ref{Eckhardt},\ref{Cressman},\ref{larkin2009},\ref{larkin2010}].
In the particular case of light tracer particles (referred to as floaters hereinafter)
in free-surface turbulence, non-uniform distribution is observed on the surface,
where floaters form clusters by accumulating along patchy and string-like
structures [\ref{larkin2009}].
Clustering occurs even if floaters have no inertia, and in the absence of
floater-floater interaction, surface tension effects, or wave motions [\ref{larkin2009}].
Differently from the case of inertial particles, in which clustering is driven by
inertia and arises when particle trajectories deviate from flow streamlines
[\ref{sm09}],
clusters are controlled by buoyancy, which forces floaters on the surface.
The physical mechanism governing buoyancy-induced clustering is closely connected
to the peculiar features of free-surface turbulence, which is characterized
by sources (resp. sinks) of fluid velocity where the fluid is moving upward
(resp. downward) [\ref{Eckhardt}].
Once at the surface, floaters follow fluid motions passively and leave
quickly the upwelling regions gathering in downwelling regions: here,
fluid can escape from the surface and sink whereas floaters can not, precisely because
of buoyancy [\ref{larkin2009}].

In a series of recent papers [\ref{Cressman},\ref{larkin2009},\ref{larkin2010}] it was shown that
floater clusters in free-surface turbulence form a compressible system that evolves towards
a fractal distribution in several large-eddy turnover times (measured at
the free-surface) and at an exponential rate. The macroscopic manifestation
of this behavior is strong depletion of floaters in large areas of the surface
and very high particle concentration along narrow string-like regions,
which are typical of scum coagulation on the surface of the sea [\ref{larkin2010}].
From a statistical viewpoint, this is reflected by a peaked probability
distribution function of particle concentration with power-law tails.
A proper description of such power-law distribution requires a clear
understanding of the mechanism by which floaters are segregated into
filamentary clusters. In this letter we examine such mechanism from a
phenomenological point of view, and we also quantify
cluster dynamics in connection
with the characteristic timescale of the surface vortices.
This analysis is of fundamental interest since it quantifies
the temporal persistency of clusters with respect to the
dominant surface flow scales, but reflects
practically towards modeling of dispersion
in many surface transport phenomena, such as
the spreading of phytoplankton, pollutants and nutrients
in oceanic flow [\ref{larkin2010}].

\section{Problem Formulation and Numerical Methodology}
The flow field is calculated by integrating incompressible
continuity and Navier-Stokes equations. In dimensionless form:
\begin{equation}
\frac{\partial u_i}{\partial x_i}=0~,~~~
\label{NS}
\frac{\partial u_i}{\partial t}=-u_j \frac{\partial u_i}{\partial x_j}+\frac{1}{Re_\tau}
\frac{\partial^2 u_i}{\partial x_j \partial x_j}-\frac{\partial p}{\partial x_i}+ \delta_{1,i}~,
\end{equation}
with $u_i$ the $i^{th}$ component of the fluid velocity,
$p$ the fluctuating kinematic pressure, $\delta_{1,i}$
the mean pressure gradient driving the flow, and 
$Re_{\tau}= h u_{\tau} / \nu$ the shear Reynolds
number based on the channel depth $h$ and the shear
velocity $u_{\tau}=\sqrt{ h |\ \delta_{1,i}|\ / \rho}$.
Eqns. (\ref{NS}) are solved directly using a pseudo-spectral method
that transforms field variables into wavenumber space,
through Fourier representations for the streamwise and spanwise
directions (using $k_x$ and $k_y$
wavenumbers respectively)
and a Chebyshev representation for the wall-normal non-homogeneous
direction (using $T_n$ coefficients). A two-level explicit
Adams-Bashfort scheme for the non-linear terms
and an implicit Crank-Nicolson method for the viscous
terms are employed for time advancement \ref{pan1995}.

Floaters motion is described by a set of ordinary differential
equations for velocity ${\mathbf{v}}_{p}$ and position
${\mathbf{x}}_{p}$ at each time step. In vector form:
\vspace{-0.3cm}
\begin{equation}
\frac{\d  {\mathbf{x}}_{p}}{\d t}={\mathbf{v}}_{p}~,~~
\frac{\d {\mathbf{v}}_{p}}{\d t}=\frac{(\rho_{p}-\rho_{f})}{\rho_{p}}{\mathbf{g}}+
\frac{({\mathbf{u}}_{@p}-{\mathbf{v}}_{p})}{\tau_{p}} f(Re_{p})
\label{pem}
\end{equation}

where ${\mathbf{u}}_{@p}$ is the fluid velocity at the
floater position, interpolated with 6th-order Lagrange
polynomials, $\rho_p$ (resp. $\rho_f$) is the
floater (resp. fluid) density, and
$\tau_{p}=\frac{\rho_{p}\, d_{p}^{2}}{18\,\rho_f \nu}$
is the floater relaxation time based on the diameter $d_{p}$.
The Stokes drag coefficient is computed using the
Schiller-Naumann non-linear correction ($f(Re_{p})=(1+0.15\, Re_{p}^{0.687})$) 
required for floater Reynolds number
$Re_{p}=|{\mathbf{u}}_{@p}- {\mathbf{v}}_{p}|\, d_{p} / \nu > 1$.
To calculate individual trajectories,
a Lagrangian tracking routine is coupled to the
flow solver [\ref{sm09}]. Periodic boundary conditions are imposed on
floaters moving outside the computational 
domain in the homogeneous directions.Eqns. (\ref{pem}) are advanced in time using a
4th-order Runge-Kutta scheme starting from a random distribution
of floaters with velocity ${\mathbf{v}}_{p}(t=0) \equiv {\mathbf{u}}_{@p}(t=0)$.

Results presented in this paper are relative to two values of the
shear Reynolds number: $Re_{\tau}^L=171$ and $Re_{\tau}^H=509$
corresponding, respectively, to shear velocity
$u_{\tau}^L=0.00605\ m s^{-1}$ and $u_{\tau}^H=0.018 \ m s^{-1}$. 
The size of the computational domain in wall units is
$L_{x}^{+} \times L_{y}^{+} \times L_{z}^{+}= 2 \pi Re_{\tau} \times \pi Re_{\tau} \times Re_{\tau}$,
discretized with $128 \times 128 \times 129$  grid points 
($k_x=i 2 \pi / L_x$, $k_y= j 2 \pi / L_y$ with $i,j=1, ..., 128$, and
$T_n(z)=\cos[n \cdot \cos^{-1}(z/h)]$ with $n=1, ..., 129$ before de-aliasing)
at $Re_{\tau}^L$ and with
$256 \times 256 \times 257$ grid points ($i,j=256$ and $n=257$
before de-aliasing) at $Re_{\tau}^H$.
Samples of ${\cal{N}}=2 \cdot 10^5$ floaters characterized by specific
density $S=\rho_p/\rho_f=0.5$ and diameter $d_p=250~\mu m$ (a
value in the size range of large phytoplankton cells [\ref{ruiz}])
were considered. The corresponding values of the non-dimensional
response time (Stokes number) $St=\tau_p/\tau_f$ with
$\tau_f=\nu / u_{\tau}^2$ the viscous timescale of the flow, are
$St^L=0.064$ at $Re_{\tau}^L$ and $St^H=0.562$ at $Re_{\tau}^H$.
Floaters with density much less than that of the fluid were
considered on purpose to confine their motion to the free
surface and produce a behavior which resembles not at all
that of neutrally buoyant, non-inertial particles.

\section{Results and discussion}
\subsection{Characterization of free-surface turbulence through energy spectra}
\begin{figure}[!]

\includegraphics[height=41.5mm, keepaspectratio, angle=-90]{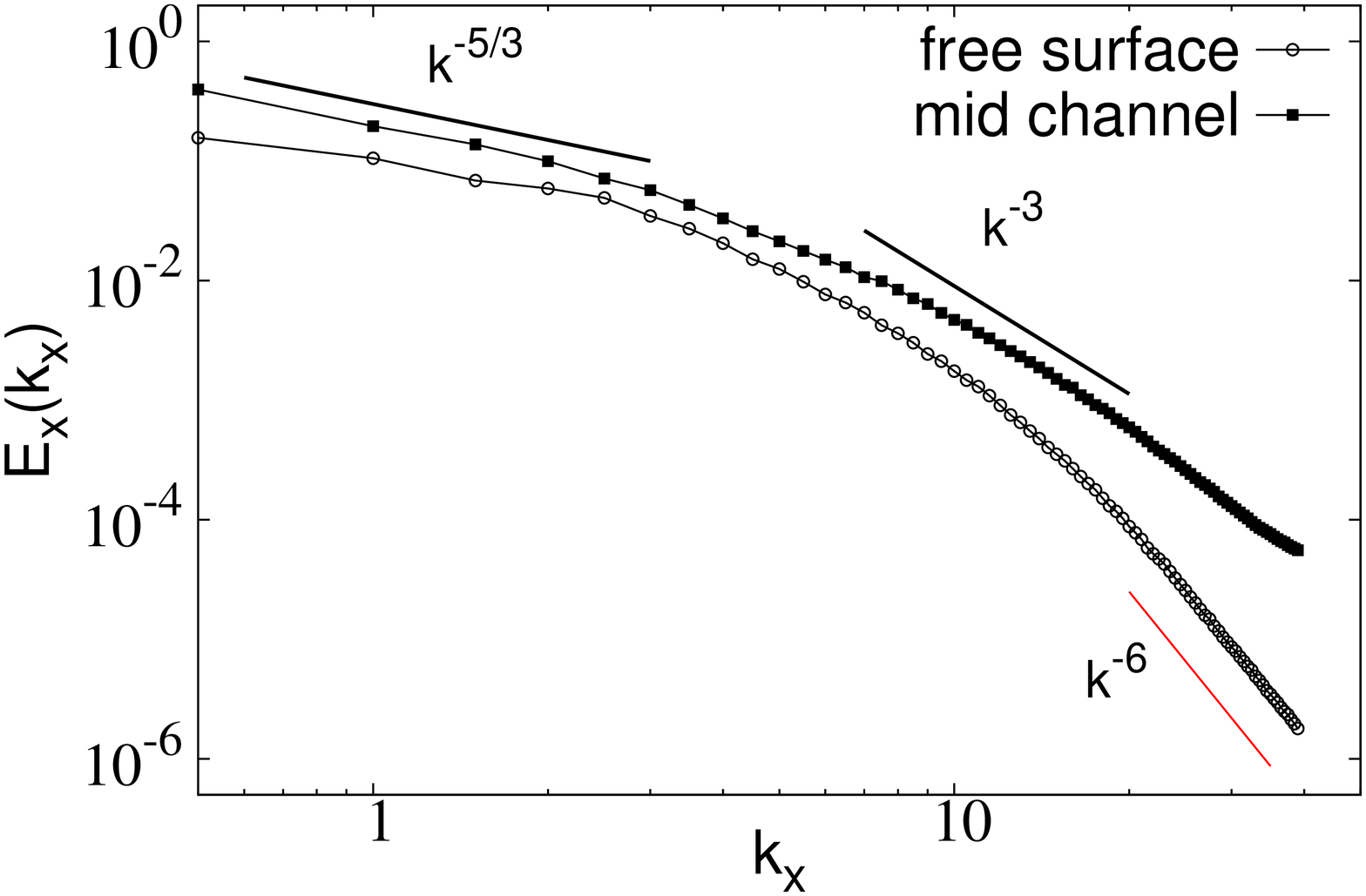}
\includegraphics[height=41.5mm, keepaspectratio, angle=-90]{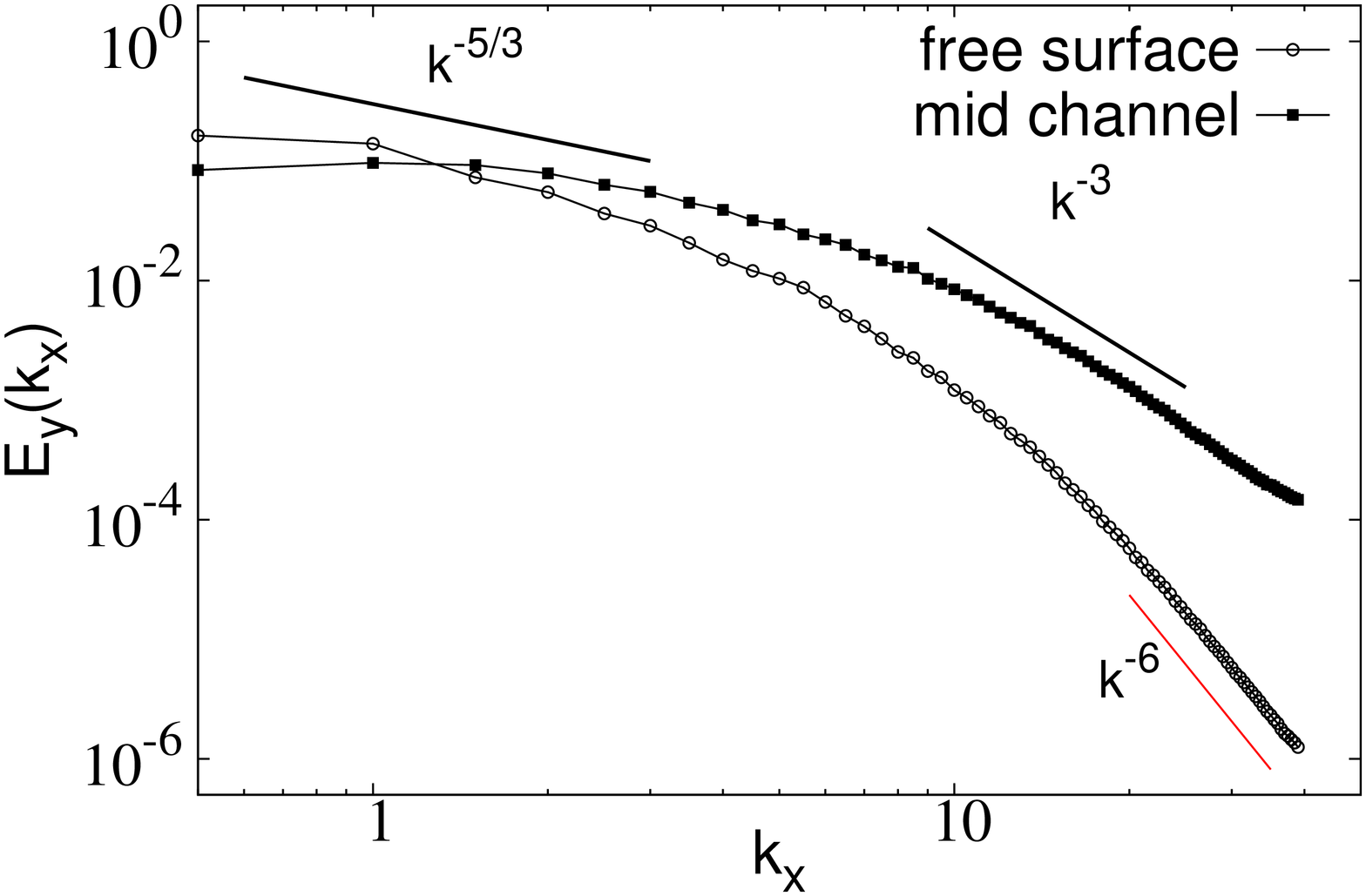}
\includegraphics[height=41.5mm, keepaspectratio, angle=-90]{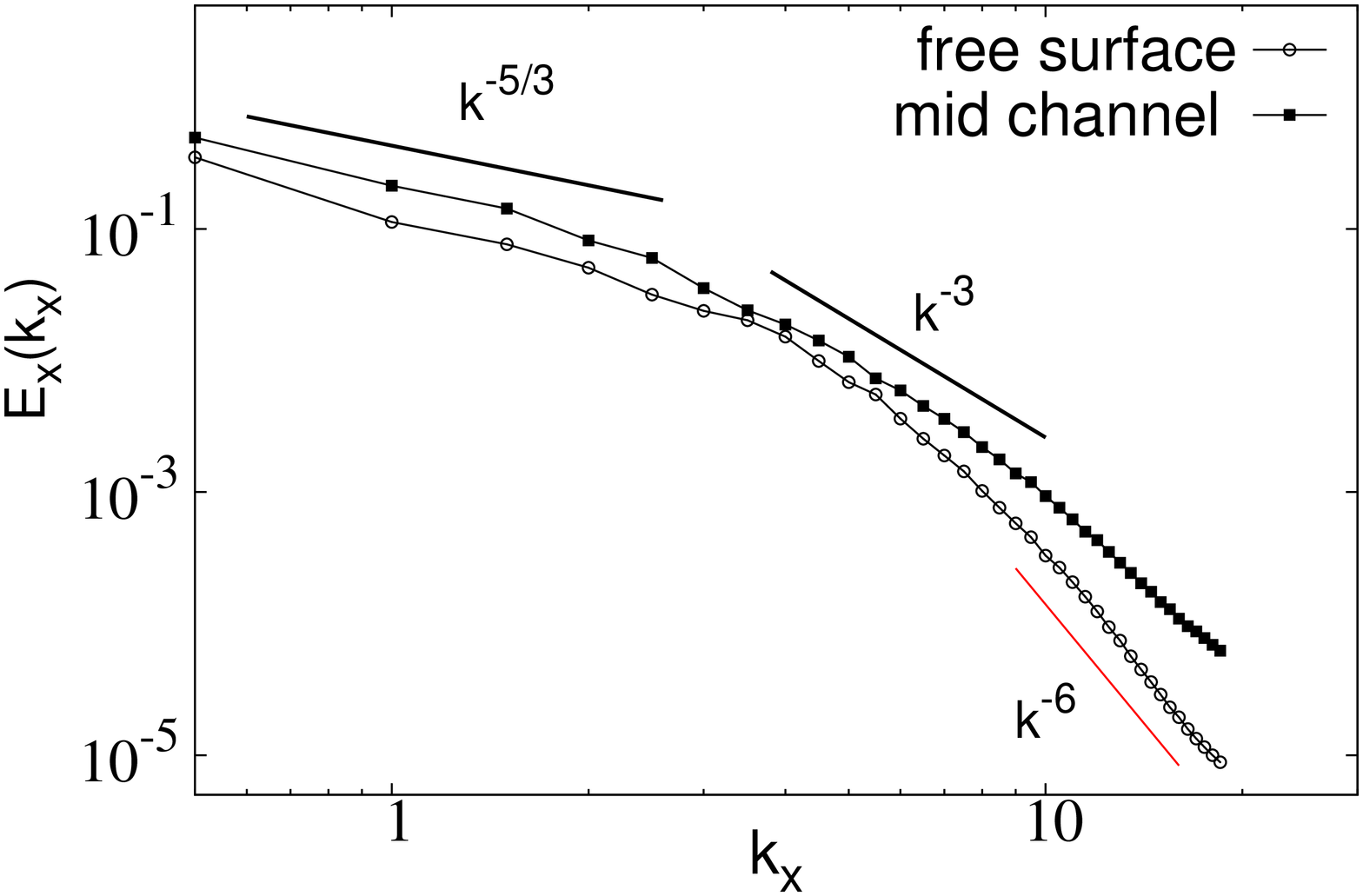}
\includegraphics[height=41.5mm, keepaspectratio, angle=-90]{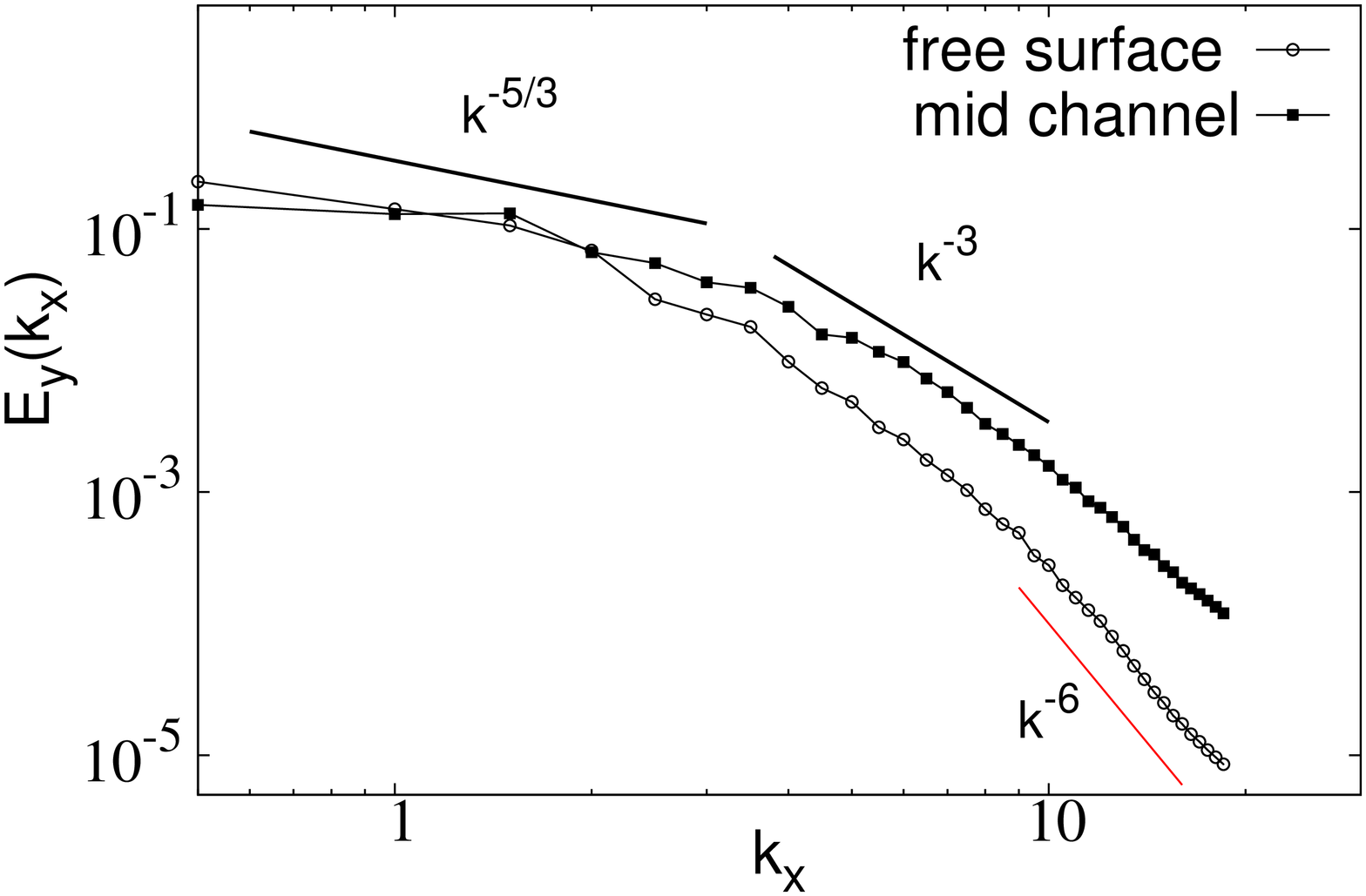}

\caption{One-dimensional (streamwise) energy spectra of the streamwise
($E_x(k_x)$, panels (a) and (c)) and spanwise ($E_y(k_x)$, panels (b)
and (d)) surface-parallel velocity fluctuations.
Spectra are computed at the free surface ($z^+=0$,
circles) and at the channel center ($z^+=254.6$ at $Re_{\tau}^H$,
$z^+=85.5$ at $Re_{\tau}^L$, squares).
}

  \label{spectra}

\vspace{-6.8cm}

\hspace{0.05cm} 
         
 (a) $Re_{\tau}^H$  \hspace{3.3cm} (b) $Re_{\tau}^H$ \hspace{1.3cm}
         
\vspace{2.2cm}

\hspace{0.05cm}

 (c) $Re_{\tau}^L$ \hspace{3.3cm} (d) $Re_{\tau}^L$ \hspace{1.3cm}
  \vspace{4.cm}

	\end{figure}

Turbulent flow structures near the free surface of an open
channel have been investigated in several previous
studies [\ref{Kermani},\ref{pan1995},\ref{Komori},\ref{Rashidi},\ref{Komori2},\ref{Nagaosa},\ref{Nagaosa12}].
All these studies show that surface structures
are generated and sustained by bursting phenomena that are continuously
produced by wall shear turbulence inside the buffer layer.
Specifically, bursts emanate from the bottom of the channel and
produce upwelling motions of fluid as they are
convected toward the free surface.
Near the surface, turbulence
is restructured and nearly two-dimensionalized due to
damping of vertical fluctuations [\ref{sarpkaya}]:
upwellings appear as two-dimensional sources for the
surface-parallel fluid velocity and alternate to sinks
associated with downdrafts of fluid from the surface to the bulk. 
Through sources fluid elements at the surface are
replaced with fluid from the bulk, giving rise to the
well-known surface-renewal events [\ref{Komori}].
Whirlpool-like vortices may also form in the high-shear region
between the edges of two closely-adjacent upwellings.
This phenomenology has been long recognized to produce flow
with properties that differ from those typical of two-dimensional
incompressible Navier-Stokes turbulence [\ref{Kumar},\ref{Eckhardt}].
These properties can be quantified examining the energy spectra of
the fluid velocity fluctuations on the surface [\ref{Nagaosa}],
shown in Fig. \ref{spectra}
for the case of statistically-steady turbulence.
To emphasize direction-related aspects of the energy spectra,
results for the surface-parallel velocities are examined in isolation:
Panels (a) and (c) in Fig. \ref{spectra} show the one-dimensional streamwise
spectra of the streamwise velocity $E_x(k_x)$ computed at the
free surface (circles) and at the channel center (squares) in the
$Re_{\tau}^L$ and $Re_{\tau}^H$ simulations, respectively;
panels (b) and (d) show the spectra of the spanwise
velocity $E_y(k_x)$ in the same two regions.
Solid lines represent the slope of the spectrum within the
inertial regimes predicted by the Kraichnan-Leith-Batchelor
phenomenology of two-dimensional turbulence [\ref{kraichnan},\ref {batchelor}]:
$k_x^{-5/3}$, representing inverse cascade of energy to large flow scales,
and $k_x^{-3}$, representing direct cascade of enstrophy to small flow scales.
A collective analysis of spectra shown in Fig. \ref{spectra}
reveals that $-5/3$ range can be observed only for few of the lowest wavenumbers.
A relatively larger range of high wavenumbers can be identified over which
spectra exhibit a $-3$ scaling. In essence, there is a sort of reverse cascade with 
energy in the high wave numbers decaying rapidly, but low wave numbers (large
vortices) decaying much more slowly.
Examining $E_x(k_x)$, we notice that the spectrum at the
free surface is always below that computed in the center of the channel.
Also, energy in the high-wavenumber portion of the spectrum decays
more rapidly [\ref{Nagaosa}], roughly as $k^{-6}$: this tendency is particularly evident at
$Re_{\tau}^H$ and indicates that only large-scale surface structures
survive to the detriment of small-scale ones.
Examining $E_y(k_x)$, we observe that redistribution
of energy from small to large scales in proximity of the free surface
determines a cross-over between spectra at low wavenumbers (for both
Reynolds numbers): this finding
confirms further that small scale structures
play little role in determining turbulence properties
in this region of the flow.

\subsection{Characterization of particle clustering through surface divergence}
\begin{figure}[!]
    \vspace{-0.cm}

\centering
\includegraphics[width=45.5mm,keepaspectratio, angle=-90]{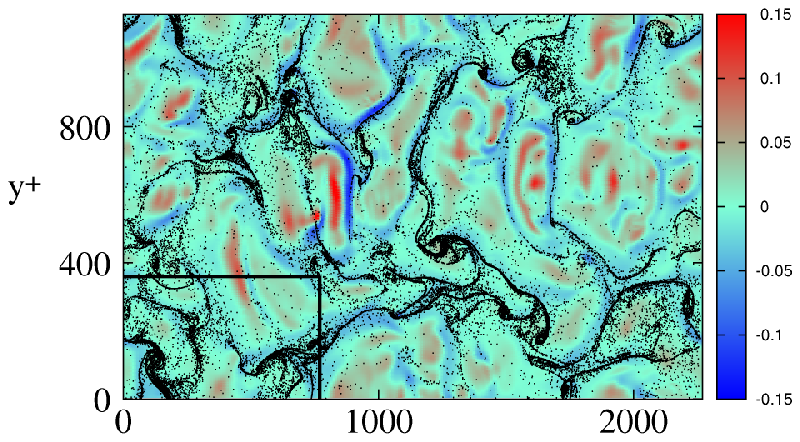}
\includegraphics[width=51.8mm, keepaspectratio, angle=-90]{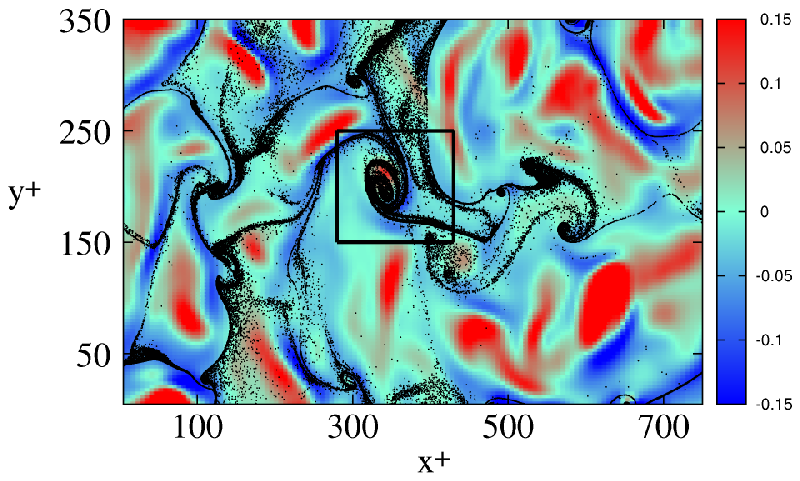}

\caption{Correlation between floater clusters and surface divergence $\nabla_{2D}$:
floaters segregate in $\nabla_{2D}<0$ regions (in blue, footprint of sub-surface
downwellings) avoiding $\nabla_{2D}>0$ regions (in red, footprint of sub-surface upwellings).
Panels: (a) $Re_{\tau}^{H}$, $t^+=180$ upon floater injection;
(b) $Re_{\tau}^{L}$, $t^+=121$. The rectangle in panel (a) renders the relative
domain size in the $Re_{\tau}^{L}$ simulation; the rectangle in panel (b) highlights
the floater cluster shown in Fig. \ref{evolution}.
}
\label{distribution}

         \vspace{-12.8cm}
         
         \hspace{-1.5cm} (a)   \hspace{6.cm}
         
         \vspace{4.1cm}
         \hspace{-1.5cm} (b)   \hspace{6.cm}
         
         \vspace{-3.8cm}
         \hspace{13.2cm} $\nabla_{2D}$ \hspace{-8cm}
         
         \vspace{3.9cm}

         \hspace{14.2cm} $\nabla_{2D}$ \hspace{-8cm}
 \vspace{6.5cm}
\end{figure}

Most of the analyses for geophysical flows were conducted
considering two-dimensional incompressible homogeneous isotropic
turbulence [\ref{dritschell},\ref{tabeling}].
In such flows the divergence of the velocity field is zero by 
construction. However, the divergence in real surface flows
is defined as:
\begin{equation}
\nabla_{2D}=\frac{\partial u}
{\partial x}+\frac{\partial v}{\partial y}=
- \frac{\partial w}{\partial z}~,
\end{equation}
and does not vanish. Therefore floaters, forced to stay on surface,
probe a compressible two-dimensional system [\ref{Cressman}], where
velocity sources are regions of local flow expansion
($\nabla_{2D} > 0$) generated by sub-surface upwellings
and velocity sinks are regions of local
compression ($\nabla_{2D} < 0$) due to downwellings [\ref{Eckhardt}].
In Fig. \ref{distribution} we provide a qualitative characterization
of floater clustering on the
free surface by correlating the instantaneous particle patterns with
the colormap of $\nabla_{2D}$.
Due to buoyancy, floaters reaching the free surface can not
retreat from it following flow motions: they can only leave
velocity sources (red areas in Fig. \ref{distribution})
and collect into velocity sinks (blue areas in Fig. \ref{distribution}).
Once trapped in these regions, floaters organize themselves in
clusters that are stretched by the fluid forming filamentary structures.
Eventually sharp patches of floater density distribution
are produced, which correlate very well with the
rapidly-changing patches of $\nabla_{2D}$, as clearly shown
by Fig. \ref{distribution}.
Similar behavior (formation of clusters with fractal
mass distribution) has been observed in previous
studies [\ref{larkin2009},\ref{Cressman}] for the case of Lagrangian
tracers in surface flow turbulence without mean shear.

\subsection{Time scaling of floaters clustering}
Due to the close phenomenological connection between clustering and 
surface turbulence, the cluster length and time scales
are expected to depend on local turbulence properties.
In particular, one can quantify the temporal coherence of surface flow
structures through their Lagrangian integral timescale (or, equivalently,
their eddy turnover time [\ref{Eckhardt}]):
\begin{equation}
T_{{\cal L},ij}=\int_0^\infty \! R_{f,ij}(t,{\mathbf{x}}_f(t)) \d t~,
\label{lagint}
\end{equation}
where:
\begin{equation}
R_{f,ij}(t,{\mathbf{x}}_f(t))=\frac{\langle {\mathbf{u'}}_{f,i}(t,{\textbf{x}}_f(t)) \cdot {\mathbf{u'}}_{f,j}(t_0,{\textbf{{x}}}_f(t_0)) \rangle}{\langle  {\mathbf{u'}}_{f,i}(t_0,{\textbf{x}}_f(t_0)) \cdot {\mathbf{u'}}_{f,j}(t_0,{\textbf{x}}_f(t_0)) \rangle}
\end{equation}
is the correlation coefficient of velocity fluctuations, obtained upon ensemble-averaging
(denoted by angle brackets) over a sample of $\cal{N}$ massless fluid tracers
released within the flow domain.
Subscript $f$ denotes the dependence of $R_{f,ij}$ on the instantaneous
position ${\mathbf{x}}_f(t)$ of fluid tracers.
Velocity fluctuations were computed as
${\mathbf{u'}}_{f,i}(t,{\mathbf{x}}_{f,i}(t))={\mathbf{u}}_{f,i}(t,{\mathbf{x}}_{f,i}(t))- {\mathbf{\bar{{u}}}}_{f,i}(t,{\mathbf{x}}_{f,i}(t)) $ ,
with ${ {\mathbf{\bar{{u}}}}_{f,i}(t,{\mathbf{x}}_{f,i}(t)) }$
the space-averaged Eulerian fluid velocity.
Estimation of $T_{{\cal L},ij}$ is crucial to parameterize particle spreading rates
and model large-scale diffusivity in bounded shear dispersion \ref{spydell}.
To compute $T_{{\cal L},ij}$ we divided the channel height into 50
uniformly-spaced bins filled with fluid tracers. For each tracer
we computed the instantaneous value of the diagonal elements
of $R_{f,ij}$
and their integral over time to get $T_{{\cal L},11}$,
$T_{{\cal L},22}$ and $T_{{\cal L},33}$. Finally, these were
ensemble-averaged within each bin using only tracers initially
located within the bin.

\begin{figure}[!]
\centering
\includegraphics[width=85mm, keepaspectratio,angle=0]{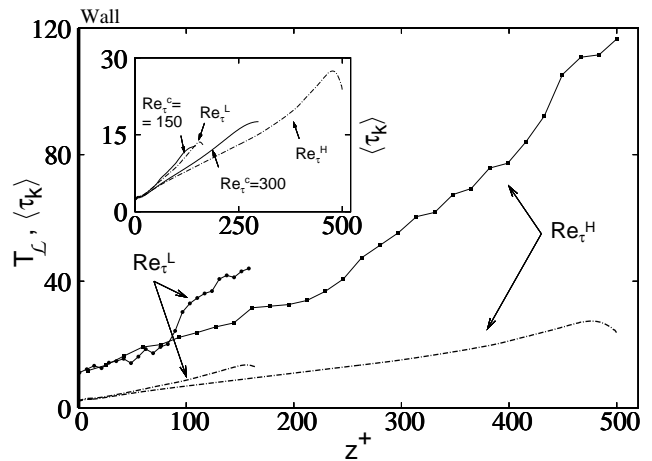}\\
\caption{Lagrangian integral fluid timescale ($T_{\cal{L}}$, symbols) and Kolmogorov
timescale ($\langle \tau_K \rangle$, lines) in open channel flow
at $Re_{\tau}^{H}$ (squares) and at $Re_{\tau}^{L}$ (circles), as function of
the wall-normal coordinate
$z^{+}$. The inset compares the behavior of $\langle \tau_K \rangle$ in open channel flow
with that in closed channel flow (at $Re_{\tau}^{c}=150$ and $300$, solid lines).
}
\label{TL}
\end{figure}

In Figure \ref{TL} we show, for both $Re_{\tau}^H$ and $Re_{\tau}^L$,
the wall-normal behavior of the Lagrangian
integral timescale of the fluid (symbols), obtained as
$T_{\cal L}=\left[ \langle T_{{\cal L},11} \rangle + \langle T_{{\cal L},22} \rangle
+ \langle T_{f,33} \rangle \right] /3$.
Note that $\langle T_{{\cal L},33} \rangle \simeq 0$ at the surface.
For comparison purposes, the Kolmogorov timescale, $\langle \tau_K \rangle$,
is also shown (dot-dashed line).
The value of $T_{{\cal L}}$ changes
significantly with the distance from the wall: in the $Re_{\tau}^H$ simulation,
$T_{{\cal L}} \simeq 120$ at the surface, a value 10 times larger than that
near the wall (where $T_{{\cal L}} \simeq 14$) indicating that the characteristic
lifetime of surface structures is significantly longer than that of
near-wall structures.
It is also evident that $T_{{\cal L}}$ is everywhere larger than
$\langle \tau_K \rangle$, confirming clear scale separation between
large-scale surface motions and small-scale dissipative structures.

\begin{figure}[]
\hspace{-0.95cm}
\includegraphics[height=43mm, keepaspectratio, angle=-90]{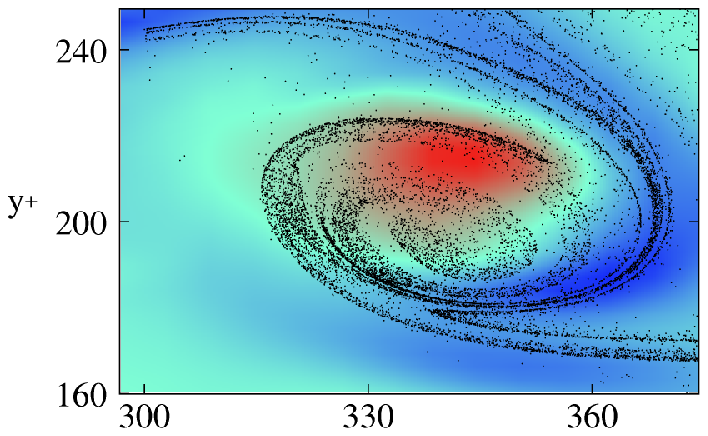}
\includegraphics[height=43mm, keepaspectratio, angle=-90]{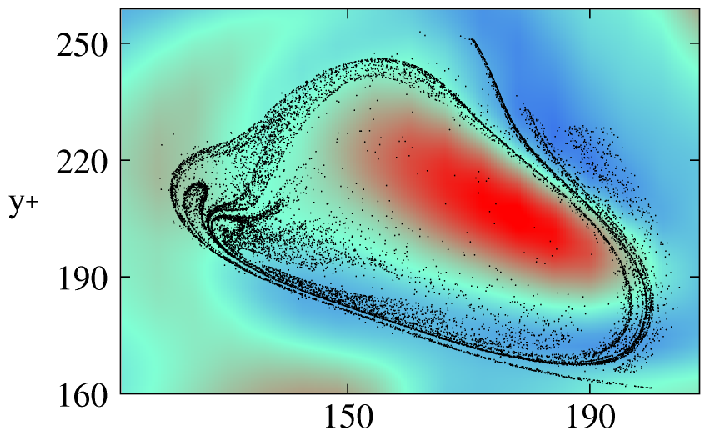}
\includegraphics[height=43mm, keepaspectratio, angle=-90]{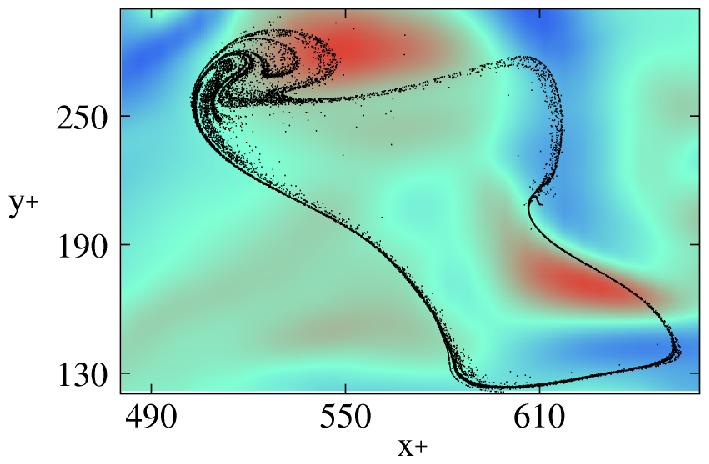}

\caption{Time evolution of the floater cluster highlighted in Fig. \ref{distribution}(b).
The cluster is examined following its Lagrangian path, with Eulerian coordinates
in each snapshot changing accordingly. Floaters are hit by an upwelling
(red region) at time $t^+ \simeq 121$ (a), and scattered around at time $t^+ \simeq 145$ (b).
Eventually, they form a highly-concentrated filamentary pattern at
time $t^+ \simeq 193$ (c). This pattern exhibits strong time persistency and over-lives
several surface-renewal events. }
\label{evolution}

         \vspace{-9.cm}

         \hspace{-3.6cm} (a)   \hspace{4.cm}  (b)    \hspace{1.4cm}
        \vspace{2.5cm}

        \hspace{-0.5cm}   (c)   \hspace{3.5cm}
    \vspace*{5.5cm}
\end{figure}

To correlate the typical lifetime of surface motions
with that of floater clusters, we examine next the
time-evolution of the local correlation dimension of clusters, $D_2(t)$ \ref{larkin2010}.
The same observable was studied experimentally also by Larkin et al.
[\ref{larkin2010},\ref{larkin2009}]
as a measure of the fractal dimension of floater distribution. 
The main finding of their analysis is that the ensemble-averaged
$D_2(t)$ decays at an exponential rate from $D_2(t=0)\simeq 2$
to $D_2(t \rightarrow \infty) \simeq 1$, the decay time being
approximately one surface eddy turnover time (defined as the typical
time for the "largest" eddies to significantly distort in
a turbulent flow).
In this work, we computed $D_2(t)$ for several surface clusters, one
of which is followed in time in Fig. \ref{evolution}.
This particular cluster was generated by past upwelling motions, which
it survived, and is now found sampling a region of the free-surface
reached by another upwelling motion (Fig. \ref{evolution}(a), red area).
Floaters are swept from the velocity source and redistribute
at its edges maintaining the cluster spatial connection, as shown in Fig. \ref{evolution}(b).
As time progresses (Fig. \ref{evolution}(c)), the cluster reshapes
generating sharp density fronts.

\begin{figure}[!]
\centering
\includegraphics[width=55mm,keepaspectratio, angle=270]{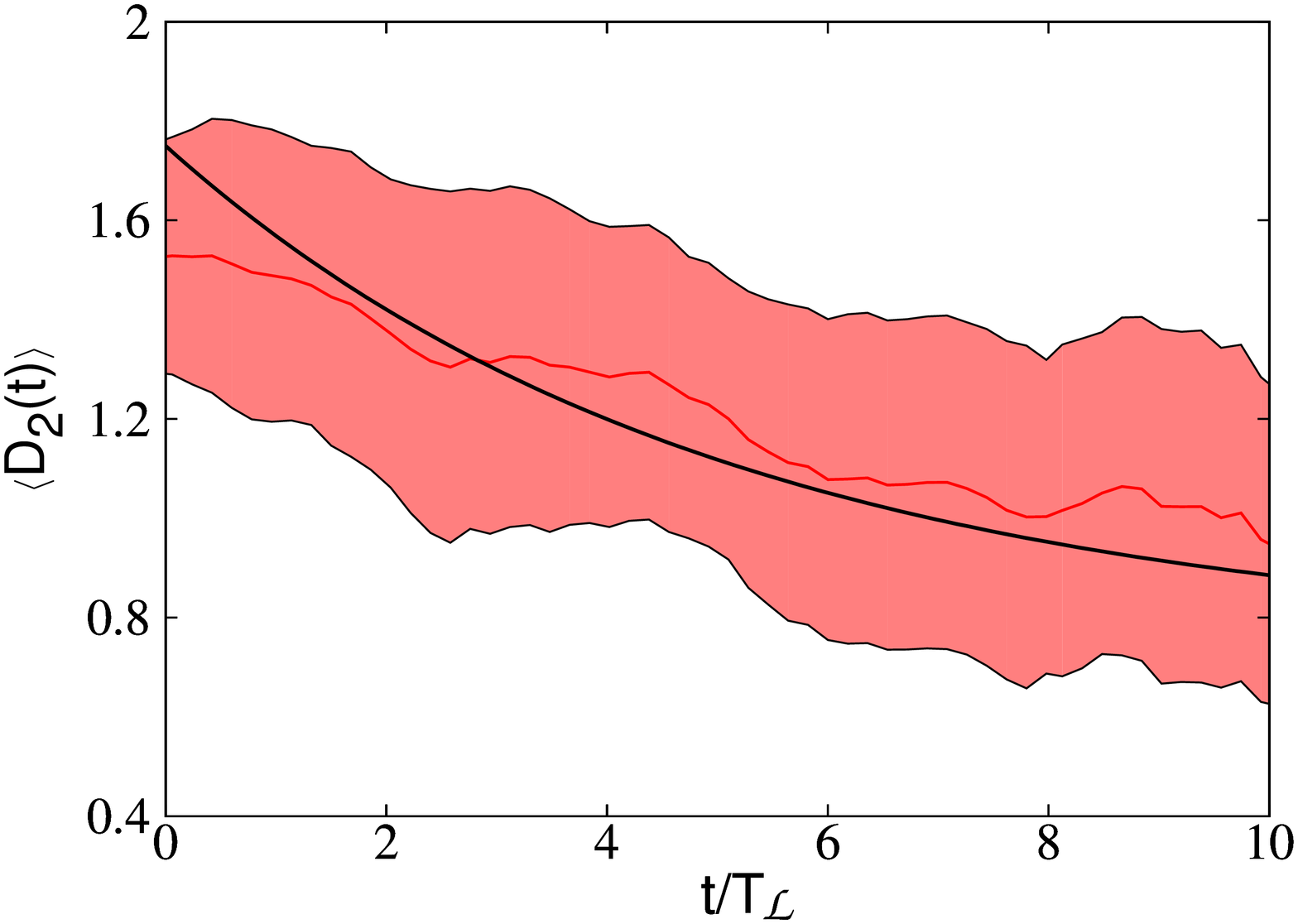}
\includegraphics[width=55mm,keepaspectratio, angle=270]{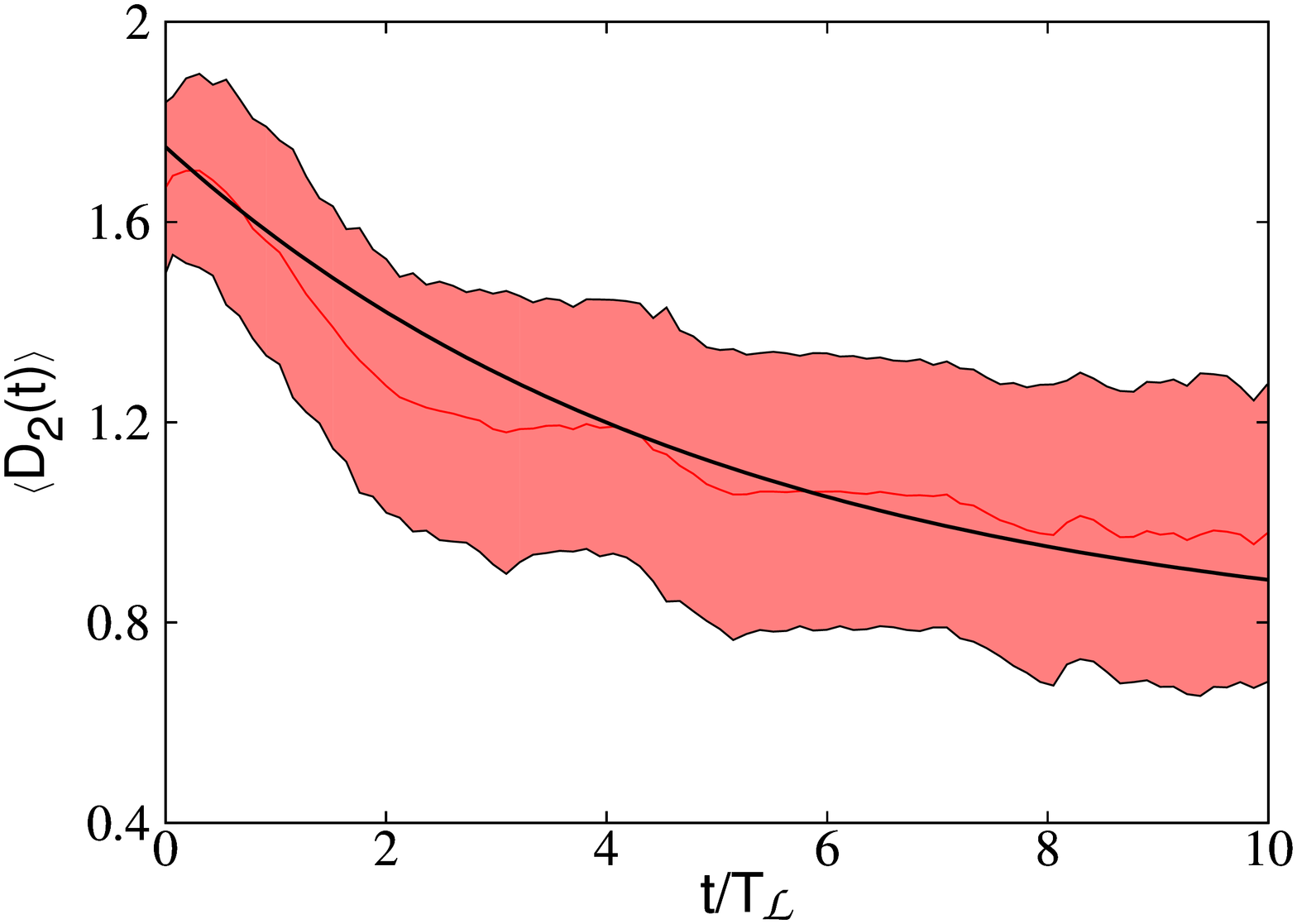}
\caption{Time evolution of the cluster correlation dimension $\langle D_2(t) \rangle$
at the free surface. Panels: (a) $Re_{\tau}^{H}$; (b) $Re_{\tau}^{L}$.
The red line in each panel shows the mean correlation dimension,$\langle D_2(t) \rangle$,
the black line represents the estimate of $\langle D_2(t) \rangle$ yield
by an exponentially-decaying fit:$\langle D_2(t) - D_2(\infty) \rangle \propto exp(-t/ \alpha T_{\cal L})$ with
$\alpha \simeq 5$ for both $Re_{\tau}^{H}$ and $Re_{\tau}^{L}$.
The red area represents the standard deviation from $\langle D_2(t) \rangle$.
Time is given as proportional to the Lagrangian integral timescale of the flow
at the free-surface.}
\label{corr-dim}

         \vspace{-10.4cm}

         \hspace{1.2cm} (a) $Re_{\tau}^{H}$  \hspace{5.5cm}

         \vspace{5.1cm}

         \hspace{1.2cm} (b) $Re_{\tau}^{L}$  \hspace{5.5cm}

         \vspace{-10.2cm}

         \hspace{20.3cm}  {\color{red} ------} $\langle D_2 (t) \rangle$  \hspace{-3.5cm}

         \vspace{-0.2cm}

         \hspace{20.3cm}  {\color{black} ------} Exp. fit  \hspace{-3.5cm}

         \vspace{-0.2cm}

         \hspace{20.3cm}  {\color{red!50} $\blacksquare$ \hspace{-0.3cm} $\blacksquare$ \hspace{-0.3cm} $\blacksquare$} $\sigma_{\langle D_2 (t) \rangle}$ \hspace{-3.5cm}

         \vspace{3.8cm}

         \hspace{20.3cm}  {\color{red} ------} $\langle D_2 (t) \rangle$  \hspace{-3.5cm}

         \vspace{-0.2cm}

         \hspace{20.3cm}  {\color{black} ------} Exp. fit  \hspace{-3.5cm}

         \vspace{-0.2cm}

        \hspace{20.3cm}  {\color{red!50} $\blacksquare$ \hspace{-0.3cm} $\blacksquare$ \hspace{-0.3cm} $\blacksquare$} $\sigma_{\langle D_2 (t) \rangle}$ \hspace{-3.5cm}
             \vspace{8.2cm}

\end{figure}

Upon isolating the floaters sub-sample $\Phi_j$ for each
cluster forming on the surface,
we computed at each time step the conditioned correlation dimension
$D_2(\Phi_j,t)$. In Fig. \ref{corr-dim}, we show the time behavior of the
ensemble-averaged correlation dimension,
$\langle D_2(t) \rangle = \sum_{j=1}^{{\cal{N}}_c} D_2(\Phi_j,t)$ (red line),
with ${\cal{N}}_c$ the number of clusters over which averaging was made
(${\cal{N}}_c=10$ for the profiles shown in Fig. \ref{corr-dim}).
To render the intermittency of the clustering phenomenon, and to quantify the
uncertainty associated with our measurement, we also plot the standard
deviation from $\langle D_2(t) \rangle$ (red area).
The black line in each panel represents the estimate of $\langle D_2(t) \rangle$
obtained assuming an exponential decay rate [\ref{larkin2010},\ref{larkin2009}].
In the present flow configuration, the decay time is given as proportional
to the value of $T_{{\cal L}}$ at the
free-surface. The best fit to the data is given by a relation of the
type $\langle D_2(t) - D_2(\infty) \rangle \propto exp(-t/ \alpha T_{\cal L})$ with
$\alpha \simeq 5$ for both $Re_{\tau}^{H}$ and $Re_{\tau}^{L}$.
This result proves the long-time persistency of  surface clusters,
that evolve in a time significantly larger than $T_{\cal L}$
to a steady state where the measured $\langle D_2(t) \rangle$ approaches a value
approximately equal to 1, in agreement with the formation of filament-like structures
observed in Fig. \ref{evolution}.
Present findings confirm qualitatively those of Larkin et al. [\ref{larkin2010},\ref{larkin2009}]
but show a slower decay time (larger than $T_{\cal L}$ and, in turn,
larger than one eddy turnover time). This may be due to the different
3D flow instance considered below the 2D free surface.

\section{Conclusions and future development}
The present study highlights the intermittent character of particle
spatial distribution in free-surface turbulence. Intermittency is due
buoyancy-driven clustering effects that are in turn
connected to the formation of sources and sinks of fluid velocity generated
by sub-surface upwelling and downwelling motions in the bulk of the 
sub-surface flow.
At small time scales, the process of cluster formation is driven by
the divergence of the flow field at the surface: Clusters evolve in time producing
fractal-like surface patterns that can be characterized by their correlation
dimension. Our results indicate that these patterns slowly relax
towards a long-term distribution with exponential decay rate,
requiring several Lagrangian
integral fluid timescales.
We remark here that, according to [\ref{Komori}], [\ref{Kermani}], the surface-renewal
timescale, which is usually employed to quantify interface scalar fluxes,
is much smaller than the Lagrangian timescale and is thus inappropriate
to quantify floater distribution dynamics.

Surface compressibility may play an important role in determining the
motion of passive tracers like pollutants and nutrients but also the
spreading rate of active ocean surfactants, such as phytoplankton [\ref{Cressman}].
Our findings provide useful indications to parameterize the relevant
timescales characterizing dispersion of such species and, therefore,
can assist in developing reliable predictive models [\ref{spydell}].


\vspace{-0.3cm}
\section*{References}
\newcounter{bean}
\begin{list}%
{\arabic{bean}.}
{\usecounter{bean}\setlength{\rightmargin}{\leftmargin}}
\item
B. Eckhardt, and J. Schumacher,
Phys. Rev. E. $\mathbf{64}$, 016314 (2001).
\label{Eckhardt}
\item
J. R. Cressman, J. Davoudi, W. I. Goldburg, and J. Schumacher,
New J. Phys. $\mathbf{6}$, 53  (2004).
\label{Cressman}
\item
J. Larkin, M. M. Bandi, A. Pumir, and W. I. Goldburg,
Phys. Rev. E $\mathbf{80}$, 066301 (2009).
\label{larkin2009}
\item
J. Larkin, W. I. Goldburg, and M. M. Bandi,
Physica D $\mathbf{239}$, 1264 (2010).
\label{larkin2010}
\item
A. Soldati, and C. Marchioli,
Int. J. Multiphase Flow $\mathbf{35}$, 827 (2009).
\label{sm09}
\item
Y. Pan, and S. Banerjee,
Phys. Fluids $\mathbf{7}$, 1649 (1995).
\label{pan1995}
\item
J. Ruiz, D. Macias, and F. Peters,
Proc. Natl. Acad. Sci. USA {\bf 101}, 17720 (2004).
\label{ruiz}
\item
S. Komori, H. Ueda, F. Ogino, and T. Mizushina,
Int. J. Heat Mass Transfer $\mathbf{25}$, 513 (1982).
\label{Komori2}
\item
M. Rashidi, and S. Banerjee,
Phys. Fluids $\mathbf{31}$, 2491 (1988).
\label{Rashidi}
\item
S. Komori, Y. Murakami, and H. Ueda,
J. Fluid Mech. $\mathbf{203}$, 103 (1989).
\label{Komori}
\item
R. Nagaosa, and R. A. Handler,
Phys. Fluids $\mathbf{15}$, 375 (2003).
\label{Nagaosa}
\item
A. Kermani, H. R. Khakpour, L. Shen, and T. Igusa,
J. Fluid Mech. $\mathbf{678}$, 379 (2011).
\label{Kermani}
\item
R. Nagaosa, and R. A. Handler,
AIChE J. $\mathbf{58}$, 3867 (2012).
\label{Nagaosa12}
\item
T. Sarpkaya,
Annu. Rev. Fluid. Mech. $\mathbf{28}$, 83 (1996).
\label{sarpkaya}
\item
S. Kumar, R. Gupta, and S. Banerjee,
Phys. Fluids $\mathbf{10}$, 437 (1998).
\label{Kumar}
\item
R. H. Kraichnan,
Phys. Fluids $\mathbf{10}$, 1417 (1967).
\label{kraichnan}
\item
G. K. Batchelor,
Phys. Fluids $\mathbf{233}$, 233 (1969).
\label{batchelor}
\item
D. G. Dritschell, and B. Legras,
Phys. Today $\mathbf{46}$, 44 (1993).
\label{dritschell}
\item
P. Tabeling,
Phys. Reports $\mathbf{362}$, 1 (2002).
\label{tabeling}
\item
M. S. Spydell, and F. Feddersen,
J. Fluid Mech. $\mathbf{691}$, 69 (2012).
\label{spydell}
\end{list}





\end{document}